\documentclass[twocolumn,prc,floatfix,showpacs,preprintnumbers,nofootinbib]{revtex4}
\usepackage{mathrsfs}
\usepackage{amssymb}
\usepackage{amsmath}
\usepackage[dvips]{graphicx}

\newcommand{\luv}{\Lambda_\mathrm{UV}}
\begin{document}

\author{T. Lappi}

\email{lappi@bnl.gov}

\affiliation{
Physics Department, Brookhaven
National Laboratory, Upton, NY 11973, USA
}

\title{
Energy density of the Glasma
}

\pacs{24.85.+p,25.75.-q,12.38.Mh}

\preprint{hep-ph/0606207}

\newcommand{\roots}{\sqrt{s}}

\newcommand{\xt}{\mathbf{x}_T}
\newcommand{\yt}{\mathbf{y}_T}
\newcommand{\pt}{{\mathbf{p}_T}}
\newcommand{\ptt}{p_T} 
\newcommand{\qt}{{\mathbf{q}_T}}
\newcommand{\kt}{{\mathbf{k}_T}}
\newcommand{\lt}{{\mathbf{l}_T}}
\newcommand{\Dt}{{\mathbf{D}_T}}
\newcommand{\At}{{\mathbf{A}_T}}
\newcommand{\nabt}{\boldsymbol{\nabla}_T}

\newcommand{\ptil}{\tilde{p}}
\newcommand{\ktil}{\tilde{k}}
\newcommand{\qtil}{\tilde{q}}

\newcommand{\emu}{\! e_\mu \!}
\newcommand{\enu}{\! e_\nu \!}

\newcommand{\ud}{\, \mathrm{d}}
\newcommand{\uc}{{\mathrm{c}}}
\newcommand{\ul}{{\mathrm{L}}}
\newcommand{\intd}{\int \!}
\newcommand{\tr}{\, \mathrm{Tr} \, }
\newcommand{\R}{\mathrm{Re}}
\newcommand{\nc}{{N_\mathrm{c}}}
\newcommand{\nf}{{N_\mathrm{F}}}
\newcommand{\half}{\frac{1}{2}}
\newcommand{\hc}{\mathrm{\ h.c.\ }}
\newcommand{\nosum}[1]{\textrm{ (no sum over } #1 )}
\newcommand{\na}{\, :\!}
\newcommand{\nb}{\!: \,}
\newcommand{\cf}{C_\mathrm{F}}
\newcommand{\ca}{C_\mathrm{A}}
\newcommand{\df}{d_\mathrm{F}}
\newcommand{\da}{d_\mathrm{A}}
\newcommand{\nr}[1]{(\ref{#1})} 
\newcommand{\dadj}{D_{\mathrm{adj}}}
\newcommand{\ra}{R_A}

\newcommand{\re}{\textcolor{red}}
\newcommand{\gr}{\textcolor{green}}
\newcommand{\bl}{\textcolor{blue}}    

\newcommand{\gev}{\ \textrm{GeV}}
\newcommand{\fm}{\ \textrm{fm}}
\newcommand{\ls}{\Lambda_\mathrm{s}}
\newcommand{\qs}{Q_\mathrm{s}}
\newcommand{\lqcd}{\Lambda_{\mathrm{QCD}}}
\newcommand{\as}{\alpha_{\mathrm{s}}}

\newcommand{\init}{\textrm{init}}
\newcommand{\final}{\textrm{final}}

\newcommand{\fig}{Fig.~}
\newcommand{\eq}{Eq.~}
\newcommand{\se}{Sec.~}
\newcommand{\eqs}{Eqs.~}

\begin{abstract}
The initial energy density produced in an ultrarelativistic heavy ion collision
can, in the color glass condensate framework,
be factorized into a product of the integrated gluon distributions
of the nuclei. Although this energy density is well defined
without any infrared cutoff besides the saturation scale, it is 
apparently logarithmically ultraviolet divergent.
We argue that this divergence is not physically meaningful
and does not affect the behavior of the system at 
any finite proper time.
\end{abstract}

\maketitle

\section{Introduction}

The matter produced at central
rapidities in a heavy ion collision is dominated by the small $x$ partons
in the wave function of the high energy nuclei. These degrees of freedom 
can, because of their high occupation numbers, be described
as a classical Weizs\"acker-Williams color field. The source for this field
is formed by the large $x$ partons, which are seen by the small
$x$ ones as classical color charges. 
The nonlinear interactions between the small $x$ gluons give rise to gluon saturation, and
the wavefunction is described by an energy (or $x$) dependent saturation scale.
This way of understanding the small $x$ wavefunction is known as the 
color glass condensate.
A model incorporating these physical features was written down
by McLerran and Venugopalan (MV) \cite{McLerran:1994ni,McLerran:1994ka,McLerran:1994vd}.

The initial transverse energy and gluon multiplicity in a collision of
two sheets of color glass in the MV model has been calculated  to all orders 
in the gluon field already some time ago
\cite{Krasnitz:1998ns,Krasnitz:1999wc,Krasnitz:2000gz,%
Krasnitz:2001qu,Lappi:2003bi,Lappi:2004sf}.
Recently there has been a renewed interest in the very early time 
behavior of these classical ``Glasma'' gluon
 fields \cite{Lappi:2006fp,Fries:2006pv},
in the context of pair production from the 
classical background field 
\cite{Gelis:2003vh,Gelis:2004jp,Lappi:2006nx,Gelis:2005pb,Fujii:2005rm,%
Fujii:2005vj,Kharzeev:2005iz,Kharzeev:2006zm}
and parity violation through the Chern-Simons charge density
of the fields \cite{Kharzeev:2001ev,Kharzeev:2004ey}.
More attention has also been paid to the 3-dimensional energy density
(instead of the energy per unit rapidity) of these field configurations
as a quantity that could be directly related to 
the initial conditions of hydrodynamical 
calculations~\cite{Hirano:2004rs,Hirano:2004rsb}.

The purpose of this note is to clarify some properties of the
initial energy density of the gauge fields in the MV model. 
We shall first, in Sec.~\ref{sec:initcond},
demonstrate that the initial energy density can be 
completely factorized into the product of the gluon distribution functions
of the colliding nuclei. Going to finite proper times, or looking
at the multiplicity, will change this factorization into a convolution
of the unintegrated gluon distributions.
We shall then, in Sec.~\ref{sec:correlator},
go on to discuss the known properties of the correlator of two pure 
gauge fields involved in the initial energy density and, in
Sec.~\ref{sec:pert}, try to understand the behavior of large $\pt$ modes
with the help of the 
lowest order perturbative solution of the classical field equations.

\section{Initial condition}\label{sec:initcond}

\begin{figure}
\begin{center}
\includegraphics[width=0.45\textwidth]{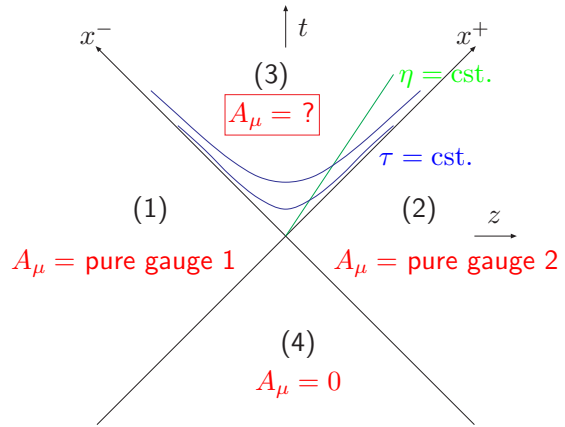}
\end{center}
\caption{The color fields in different regions of spacetime.
In regions (1) and (2), where only one of the nuclei has passed by, the field is the pure gauge field of this one nucleus. Inside region (3) the field is known numerically, but the initial condition for the field on the light cone
$\tau=0$ is determined by the two pure gauge fields.
}
\label{fig:spacet}
\end{figure}

\begin{figure}
\begin{center}
\includegraphics[width=0.45\textwidth]{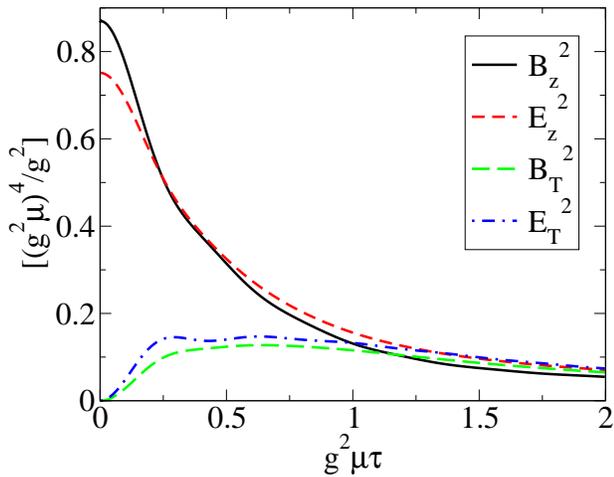}
\end{center}
\caption{
Components of the color field, computed on a $512^2$ lattice with 
$g^2 \mu \ra = 67.7$.}
\label{fig:components}
\end{figure}

Because of their high speed and Lorentz time dilation, the large $x$ degrees of 
freedom are seen by the low $x$ fields as slowly evolving in light cone time.
They form classical, static (in light cone time) sources on the light cones:
\begin{equation}\label{eq:current}
J^{\mu} =\delta^{\mu +}\rho_{(1)}(\xt,x^-)
+ \delta^{\mu -}\rho_{(2)}(\xt,x^+),
\end{equation}
where the support of the sources around the light cone must be understood 
as being very close to a delta function: 
$\rho_{(1,2)}(\xt,x^\pm) \sim \delta(x^\pm)$.
We shall work here in the Schwinger gauge 
$ A_\tau = \left(x^+ A^- + x^- A^+\right)/\tau =0$,
in which the current \nr{eq:current} is not rotated by the soft classical field;
in more general gauges \eq\nr{eq:current} should be dressed by Wilson lines
to maintain its covariant conservation.
The Weizs\"acker-Williams fields describing the softer degrees of freedom 
can then be computed from the classical Yang-Mills equation
\begin{equation}
[D_{\mu},F^{\mu \nu}] = J^{\nu}.
\end{equation}
In the light cone gauge the field of one nucleus is a pure gauge outside
the light cone (see Fig.~\ref{fig:spacet})
\begin{equation}\label{eq:pureg}
A^i_{(1,2)}(\xt) = \frac{i}{g} U_{(1,2)}(\xt) \partial_i U^\dag_{(1,2)}(\xt),
\end{equation}
where the SU(3) matrices $U_{(m)}(\xt)$ are determined
from the color sources as
\begin{equation}\label{eq:wline}
U_{(1,2)}(\xt) = 
P \exp \left\{-i g \int \ud x^\pm \frac{1}{\nabt^2} \rho_{(1,2)}(\xt,x^\pm)\right\}.
\end{equation}
In the original MV model, which we shall be using here, the color charge
densities are stochastic Gaussian random variables on the transverse 
plane: $\rho^a_{(1)}(\xt,x^-) = \delta(x^-)\rho^a(\xt) $ with
\begin{equation}\label{eq:rhorho}
\langle \rho^a(\xt) \rho^b(\yt) \rangle 
= g^2 \mu^2 \delta^{ab}\delta^2(\xt-\yt),
\end{equation}
where the density of color charges $g^2\mu$ is, up to a numerical 
constant and a logarithmic uncertainty, related to the
saturation scale $\qs$.

The initial conditions for the fields in the future light cone between the
two colliding sheets were derived and the equations of motion solved
to lowest order in the fields in 
\cite{Kovner:1995ja,Kovner:1995ts,Gyulassy:1997vt} 
(see also Ref.~\cite{Kovchegov:1997ke} for the same calculation in covariant gauge
and Ref.~\cite{Fries:2006pv} for another formulation.) 
This initial condition has a simple expression in terms
of the pure gauge fields of the two colliding nuclei \nr{eq:pureg}:
\begin{eqnarray} \label{eq:trinitc}
A^i &=& A^i_{(1)} + A^i_{(1)} \\
\label{eq:linitc}
A^\eta &=& \frac{ig}{2} \left[ A^i_{(1)} , A^i_{(2)} \right]
\\
\partial_\tau A_i &= &0
\\
\partial_\tau A^\eta &=&0.
\end{eqnarray}
Note that the metric in the $(\tau,\eta,\xt)$ coordinate system is
$g_{\mu \nu} = \mathrm{diag}(1,-\tau^2,-1,-1)$ so that
$A_\eta = -\tau^2 A^\eta$. In the Schwinger
gauge $A^\tau  = 0$ the $\pm$ components of the gauge field are related by
$A^\pm = \pm x^\pm A^\eta$.
Because of the explicit time dependence in the metric $A^\eta$
corresponds, at $\tau=0$, to the $z$-component of the 
chromoelectric field.
At the initial time the only nonzero components of the field strength
tensor are the longitudinal electric and magnetic fields and 
consequently the energy density is given by
\begin{multline}\label{eq:eps00}
\varepsilon(\tau=0) = \lim_{\tau \to 0^+} 
\frac{1}{\tau}\frac{\ud E}{\ud^2 \xt \ud \eta}
= 
\\
\half \tr F_{ij}F_{ij} + 4 \tr \left(A^\eta\right)^2.
\end{multline}

Let us introduce a shorter notation for the correlation function of
the pure gauge field of the nucleus when averaged with the distribution
\nr{eq:rhorho}. We shall define the correlation function $G(\pt)$ by
\begin{multline}\label{eq:AAcorr}
\left\langle A_i^{(m)a}(\pt) A_j^{(n)b}(\qt) \right\rangle
\equiv
\\
 (2 \pi)^2 
\delta^{mn}
\delta^{ab} \delta^2(\pt + \qt)
\frac{p_i p_j}{\pt^2} G(\pt).
\end{multline}
The index in parentheses $(m)$ refers to the two colliding nuclei, which are,
naturally, independent of each other, thus the $\delta^{mn}$
in the correlator. The correlator must also be diagonal in the
color index ($\delta^{ab}$) because there is no preferred direction
in color space present in the problem. We are assuming translational
invariance on the transverse plane ($\delta^2(\pt + \qt)$), which is justified
because we are only interested in momentum scales much larger than the
nuclear geometry effects which break this invariance (meaning that we are assuming $|\pt| \gg 1/\ra$). The transverse spatial index structure
$p_i p_j$ is the only one consistent with rotational invariance
on the transverse plane (again, at momentum scales $|\pt| \gg 1/\ra$
there is no preferred direction in the system to break this invariance).

Using the notation of \eq~\nr{eq:AAcorr} the two terms in the energy density
\nr{eq:eps00} become
\begin{widetext}
\begin{eqnarray}
\label{eq:bsqr}
\int \ud^2 \xt
\half \tr F_{ij}F_{ij} &=& \half g^2 \nc \left(\nc^2-1\right)
\pi \ra^2 \int \frac{\ud^2\kt}{(2\pi)^2}\frac{\ud^2\pt}{(2\pi)^2}
\left[\pt^2 \kt^2 - (\pt \cdot \kt)^2 \right] 
\frac{G(\pt)}{\pt^2} \frac{G(\kt)}{\kt^2}
\\
\label{eq:esqr}
\int \ud^2 \xt
4 \tr \left(A^\eta\right)^2 &=&
\half g^2 \nc \left(\nc^2-1\right)
\pi \ra^2 \int \frac{\ud^2\kt}{(2\pi)^2}\frac{\ud^2\pt}{(2\pi)^2}
\left[(\pt \cdot \kt)^2 \right] 
\frac{G(\pt)}{\pt^2} \frac{G(\kt)}{\kt^2}
\end{eqnarray}
\end{widetext}
and the final result factorizes into
\begin{equation}\label{eq:eps0}
\varepsilon|_{\tau=0} =
\frac{g^2}{2} \nc \left(\nc^2-1\right)
\left[ \int \frac{\ud^2\pt}{(2\pi)^2}
 G(\pt) \right]^2.
\end{equation}
Equation~\nr{eq:eps0} is our main result. The initial energy density factorizes 
completely into a product of two terms, both of which only depend on the 
properties of one single nucleus. This happens only strictly at $\tau=0$.

Note that due to rotational invariance
the initial energy density in the magnetic field, \eq\nr{eq:bsqr} and the 
electric field, \eq\nr{eq:esqr} are equal. The discretized version
of the computation on a transverse lattice breaks this rotational 
invariance (for an explicit expression
see the lattice perturbation theory result
in Appendix B of Ref.~\cite{Krasnitz:1998ns}).
This violation  is largest for 
the momentum modes near the edges of the Brillouin zone. As we shall 
discuss in the following, for larger proper times these modes do not
affect the energy density any more, and the energy densities
 in the longitudinal electric and magnetic fields approach each other,
as can be seen in Fig.~\ref{fig:components}.

\section{Properties of the gauge field correlator}\label{sec:correlator}

Let us then recall some known properties of $G(\pt)$, the correlation function
of the pure gauge fields defined in \eq\nr{eq:AAcorr}. In light cone 
quantization it is related to the unintegrated gluon distribution 
function\footnote{The notation and numerical constants at this point 
are very confusing. Here an attempt is made to follow 
Ref.~\cite{Iancu:2000hn} (in particular Sec.~2.4) and 
Ref.~\cite{McLerran:1994ni}.}
\begin{equation}\label{eq:gdist}
x G(x,Q^2) = \ra^2 \left(\nc^2-1\right) 
\int^{Q^2} \frac{\ud^2 \kt}{(2\pi)^2} 
G(\kt).
\end{equation}
Note, however, that our $G(\pt)$ is equivalent to the unintegrated
gluon distribution used to compute gluon production in pA-collisions
\emph{only} in the weak field limit. We refer the reader to
Refs.~\cite{Kharzeev:2003wz,Blaizot:2004wu,Gelis:2006tb}
for a discussion of the difference. Our 
$G(\pt)$ is equivalent, up to the
normalization, to $\phi^\textrm{WW}$ of Ref.~\cite{Kharzeev:2003wz}.

The correlator $G(\pt)$ has been analytically evaluated in several 
papers \cite{Jalilian-Marian:1997xn,Kovchegov:1996ty,Kovchegov:1997pc}.
The result is expressed in closed form in coordinate space as
\begin{equation}\label{eq:jalimari}
G(\xt) = \frac{4 }{g^2 \nc \xt^2}
\left(1 - e^{- \frac{\nc}{8 \pi} 
\xt^2 (g^2 \mu)^2 \ln \frac{1}{\Lambda |\xt|}} \right).
\end{equation}
Here $\Lambda$ is an infrared cutoff that one must introduce in order
to invert the 2 dimensional Laplace operator.
The same function can also be measured in the numerical setup used
to compute the glasma fields \cite{Krasnitz:1998ns,Krasnitz:1999wc,%
Krasnitz:2000gz,Krasnitz:2001qu,Lappi:2003bi}.
The numerical procedure used is not exactly equivalent
to the calculation leading to \eq\nr{eq:jalimari},
because the source in \eq\nr{eq:wline} is taken as exactly
a delta function on the light cone and the infrared divergence in 
inverting the Laplace operator $\nabt^2$ is effectively regulated 
by the size of the lattice.
These differences can, however, be absorbed into
the infrared cutoff $\Lambda$, and the numerical evaluation 
(see in particular Fig.~3 of Ref.~\cite{Krasnitz:2002mn}) of the 
correlator agrees with the behavior of \eq\nr{eq:jalimari}.
Note that to derive the correct initial conditions,
\eqs\nr{eq:pureg},~\nr{eq:wline},~\nr{eq:trinitc}
and~\nr{eq:linitc}, it is essential to consider the source 
as spread out in the longitudinal coordinate. Only when this is done
can one, in practice, take the source as a delta function on the light 
cone when evaluating the Wilson line, \eq\nr{eq:wline}.

Let us then estimate the behavior of the correlator $G(\pt)$ in momentum
space. For small momenta $G(\pt)$ diverges, but the divergence is only
logarithmic and thus integrable. 
This is the essential feature of gluon saturation; bulk quantities that are 
sensitive to the harder modes in the spectrum, such as the energy density,
are infrared finite when the nonlinear interactions are taken into account 
fully.

For large momenta $G(\pt)$ has a perturbative tail behaving as 
$1/\pt^2$, meaning that the integral $\int \ud^2 \pt G(\pt)$
and thus the initial energy density are seemingly ultraviolet divergent.
This can be seen equivalently as the logarithmic
ultraviolet divergence in 
\eq\nr{eq:jalimari},
\begin{multline}\label{eq:uvdiv}
 \int \frac{\ud^2\pt}{(2\pi)^2} G(\pt)
= \lim_{|\xt| \to 0} G(\xt) 
\\
= 
 \frac{1}{2\pi g^2} (g^2\mu)^2 \lim_{|\xt| \to 0} \ln\frac{1}{|\xt|\Lambda}.
\end{multline}
We must emphasize that although \eq\nr{eq:uvdiv} involves, for dimensional
reasons, the infrared cutoff $\Lambda$, it corresponds to a divergence
from large transverse momentum, or small distance, modes. 

\begin{figure}
\begin{center}
\includegraphics[width=0.45\textwidth]{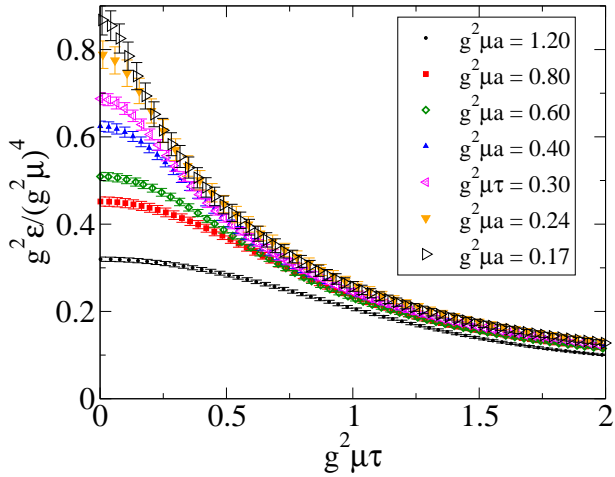}
\end{center}
\caption{The energy density for different lattice spacings
with $g^2\mu\ra = 67.7$ The continuum limit corresponds to 
$g^\mu a \to 0$. The fields configurations have been calculated
on transverse lattices of $100^2$ to $700^2$ points. 
The errors are statistical from between 20 configurations for
the smaller lattices to 5 configurations on the largest ones. }
\label{fig:totendens}
\end{figure}

The initial energy density of the glasma is infrared finite, but seemingly
logarithmically ultraviolet divergent. There are two reasons
why this divergence is fundamentally not a problem for the physical
picture of the glasma. The first reason is that, as can be seen from 
\eq\nr{eq:gdist}, the divergence corresponds to large $Q^2$
and thus, for a fixed energy, large $x$ modes in the wavefunction.
These are degrees of freedom that were, by our initial assumptions,
not meant to be included in the classical field in the first place.
It would therefore be physically well motivated to regulate
them with an ultraviolet cutoff $\luv \gtrsim \qs$, and then 
match this cutoff with whatever way one treats these hard collisions. 
This is indeed the approach advocated e.g. in Ref.~\cite{Fries:2006pv}.
The other reason for not worrying about the ultraviolet divergence is that,
as we shall argue in the following, the energy density 
of the system at later proper times $\tau \sim 1/\qs$ has a finite
$\luv \to \infty$ limit.
Thus if one regulates the ultraviolet divergence in any convenient 
way and proceeds to solve the equations forward in time, the cutoff no longer
significantly influences the later time evolution of the glasma fields.

\section{Perturbative comparison and the apparent UV divergence}\label{sec:pert}

\begin{figure}
\begin{center}
\includegraphics[width=0.45\textwidth]{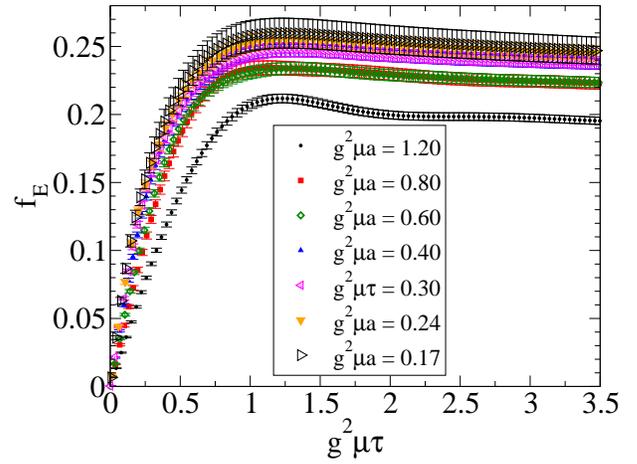}
\end{center}
\caption{The energy per unit rapidity 
$f_E \equiv \frac{g^2}{\pi\ra^2 (g^2\mu)^3} \frac{\ud E}{\ud \eta}$
(energy density divided by proper time) for different lattice spacings.
The data are from the same numerical calculation as 
Fig.~\protect\ref{fig:totendens}.}
\label{fig:toten}
\end{figure}

\begin{figure}
\begin{center}
\includegraphics[width=0.45\textwidth]{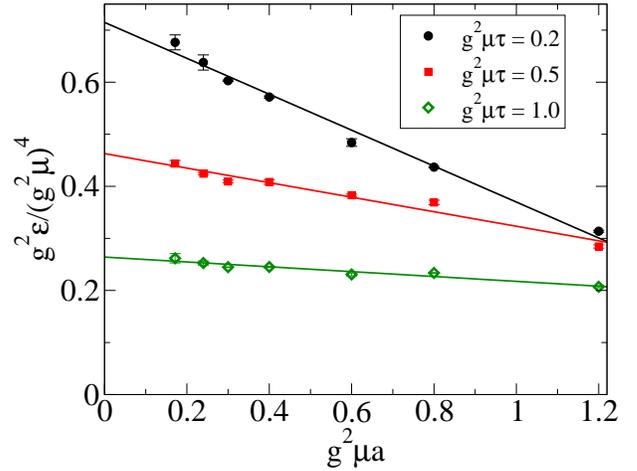}
\end{center}
\caption{The energy density at proper times
$\tau$ of 0.2, 0.5 and 1.0 in units of $1/g^2\mu$ 
as a function of the lattice spacing $a$, also
in units of $1/g^2\mu$. The data are from the same 
calculation as Fig.~\protect\ref{fig:totendens}.
One sees that the energy density converges to the
continuum limit ($g^2\mu a \to 0$) better at later times.
The straight lines are linear extrapolations.}
\label{fig:endfixtime}
\end{figure}

In the perturbative (lowest order in the source charge densities) solution
\cite{Kovner:1995ja,Kovner:1995ts,Gyulassy:1997vt,Kovchegov:1997ke}
the field amplitudes behave, in the two dimensional Coulomb gauge,
like Bessel functions
\begin{eqnarray}\label{eq:pertAi}
A_i(\tau,\kt) &=& A_i(\tau=0,\kt)J_0(|\kt|\tau) 
\\ \label{eq:pertAeta}
A_\eta(\tau,\kt) &=& \frac{2 \tau}{|\kt|} A^\eta(\tau=0,\kt)J_1(|\kt|\tau).
\end{eqnarray}
The energy density corresponding to this perturbative solution
(this time dependence is also derived in 
Refs.~\cite{Kovchegov:2005ss,Kovchegov:2005kn}) is 
\begin{multline}
\label{eq:epsbessel}
\varepsilon = \half g^2  \nc \left(\nc^2-1\right)
\int \frac{\ud^2\kt}{(2\pi)^2}\frac{\ud^2\pt}{(2\pi)^2}
G(\pt) G(\kt) 
\\ 
\times
\left[ J^2_0(|\pt+\kt|\tau)+J^2_1(|\pt+\kt|\tau)\right].
\end{multline}
To lowest order in the sources the pure gauge field correlator is
\begin{equation}
G(\pt) = \frac{1}{g^2}\frac{(g^2\mu)^2}{\pt^2}
\end{equation}
and, using the asymptotic behavior of the Bessel functions in
\eq\nr{eq:epsbessel}, the energy density for large times
reduces to 
\begin{equation}
\varepsilon(\tau\to \infty) =
\frac{1}{\tau} \int \ud^2 \kt \frac{\ud N}{\ud y \ud^2\kt} |\kt|
\end{equation}
with the Bertsch-Gunion \cite{Gunion:1981qs} type
multiplicity resulting from the lowest order solution
\cite{Kovner:1995ja,Kovner:1995ts,Gyulassy:1997vt,Kovchegov:1997ke}:
\begin{multline}
\frac{\ud N}{\ud y \ud^2 \kt} =   \frac{\pi \ra^2}{(2 \pi)^2}
\frac{\nc (\nc^2-1) g^6 \mu^4}{\kt^2} \frac{1}{\pi}
\\ \times
\int \frac{\ud^2 \pt}{(2 \pi)^2} \frac{1}{\pt^2 (\kt-\pt)^2}.
\end{multline}

Also in the full nonperturbative solution the most ultraviolet modes
behave in the same way 
\cite{Krasnitz:2001qu,Lappi:2003bi,Krasnitz:2003jw}. 
Let us now assume that we have 
computed the initial energy density with some ultraviolet cutoff 
$\luv$
(such as the inverse lattice spacing in the numerical calculation).
The initial energy density \eq\nr{eq:eps0} depends on this 
cutoff as $\ln^2 \luv$, as can be seen in the numerical
result in Fig.~\ref{fig:totendens}. After a time
$\tau \gtrsim 1/\luv$, however, the time dependence
of the modes near the cutoff changes from  the initial
$A_i(\tau) = A_i(0) \left(1 + \mathcal{O}(\tau^2) \right)$
to  the asymptotic regime 
$A_i \sim 1/|\kt| \tau$ and the contribution of these modes
to the total energy density is suppressed by an additional power of the
momentum, making the energy density finite in the limit $\luv \to \infty$.
After a time $\tau \gtrsim 1/\luv$,
or $\tau \gtrsim a$ in the lattice computation, the energy density then
converges to a value that is independent of the lattice spacing, 
as shown in Fig.~\ref{fig:toten}. The larger the proper time that one is
looking at, the better the convergence in the continuum limit is.
Figure~\ref{fig:endfixtime} shows the continuum extrapolations
of the energy densities at $g^2\mu\tau=0.2$, $0.5$ and $1.0$.

By turning this argument around one can see that if one
takes first the continuum limit 
($\luv \to \infty$ or $a\to0$) for a finite $\tau$, 
the continuum extrapolated energy density will behave as $\ln^2 1/\tau$. 
Thus if the limits are taken in this order, the energy density is indeed
finite for all $\tau>0$, but diverges logarithmically at $\tau=0$. Note
that this divergence is so weak that the energy per unit rapidity 
($\tau \varepsilon$) is still zero at $\tau=0$.
Thus we se that the solution of the field equations is well defined as an initial value problem only in the presence of an ultraviolet cutoff in the 
transverse momenta, but for later times the system loses memory of this cutoff.
Incidentally, because of this feature one can argue
that introducing of a finite initial time 
(such as done in Refs.~\cite{Romatschke:2005pm,Romatschke:2006nk}
to avoid the singularity resulting from broken boost invariance
of the field configurations) does not really add another physical
parameter into the model.

\section{Energy density in physical units and conclusion}\label{sec:conclusion}

For concreteness let us finally try to express the results shown in 
Figs.~\ref{fig:totendens},~\ref{fig:toten} and~\ref{fig:endfixtime}
in physical units. Due to the difficulty in fixing 
exactly the right value of the color charge density parameter
$g^2\mu$ this is not necessarily straightforward.
For RHIC energies one can argue, based on both 
the gluon multiplicity found in the numerical 
calculation~\cite{Krasnitz:2000gz,Krasnitz:2001qu,Lappi:2003bi}
and counting the number of large $x$ degrees of sources in the wavefunction
from conventional parton distribution functions~\cite{Gyulassy:1997vt}
that the relevant value would be $g^2\mu=2 \gev$. 
Other estimates, e.g.~\cite{Gelis:2005pb}, give a smaller value, 
so $g^2\mu=2 \gev$ should be considered an upper bound for RHIC.
For LHC energies the estimate based on parton 
distributions~\cite{Gyulassy:1997vt} gives $g^2\mu \sim 4\gev$,
but this is most certainly an overestimate, since the calculation
in Ref.~\cite{Gyulassy:1997vt} is based on parton distributions
in the proton and does not take into account
shadowing corrections. Another way of estimating the color charge density
is based on the small $x$ scaling properties
observed in deep inelastic scattering 
data~\cite{Golec-Biernat:1998js,Golec-Biernat:1999qd, Stasto:2000er,%
Freund:2002ux,Kharzeev:2004if}. The saturation scale
in this scaling can then be related to the MV model color charge 
density~\cite{Iancu:2003xm,Weigert:2005us}.
This line of thought leads to
a scaling $\qs^2 \sim (g^2\mu)^2 \sim \sqrt{s}^{\bar{\lambda}}$,
where $\bar{\lambda}  = \lambda/(1+\lambda/2)$ 
and a fit to the HERA data~\cite{Golec-Biernat:1998js} gives
$\lambda = 0.288$ and thus $\bar{\lambda}  \approx 0.25$.
The result for the color charge density 
at the LHC would be $g^2\mu \approx 3\gev,$ which is the value we will 
use in the following.

As we have seen, the energy density strictly
at $\tau=0$ is not the best quantity to look at. Let us instead
estimate the energy density at the time $\tau = 1/g^2\mu$. This is
when, as can be seen from Fig.~\ref{fig:toten}, the $1/\tau$--decrease
of the energy density seems to start. The simple
linear continuum extrapolation of Fig.~\ref{fig:endfixtime} yields
$\varepsilon(\tau=1/g^2\mu) = 0.26 (g^2 \mu)^4/g^2$. 
Using $g=2$ we then get the estimates
$\varepsilon(\tau = 0.1 \fm) \approx 130 \gev/\!\fm^3$ for RHIC 
and $\varepsilon(\tau = 0.07 \fm) \approx 700 \gev/\!\!\fm^3$
for the LHC. This estimate agrees with the values
given in Ref.~\cite{Krasnitz:2003jw} for $\tau=3/g^2\mu$
when the $\sim 1/\tau$ behavior of the energy density is taken into account.
The uncertainty due to the unprecise value of 
$g^2\mu$ in these numbers is quite large because
of the power law dependence $\varepsilon \sim (g^2\mu)^4$.

In conclusion, we have shown that the initial 3~dimensional energy density 
of the Glasma 
fields in the MV model can be expressed as a product of of the
(integrated) gluon distribution functions of the colliding nuclei.
Only the energy density at later times and the multiplicity 
involve convolutions of the unintegrated gluon distributions,
probing the $\kt$-distributions in the wavefunctions of the colliding
nuclei in more detail.
We have recalled the known properties if the pure gauge field
correlator \eq\nr{eq:AAcorr} appearing in the initial energy density. 
As expected from general gluon saturation arguments, the initial energy 
density is infrared finite when the gluon fields are solved to all orders
in the source. The energy density strictly at $\tau=0$ is, however, 
ultraviolet divergent in the MV model. We show, both by a direct numerical 
calculation and by examining the time dependence of the ultraviolet modes,
that this divergence does not persist when the equations of motion
are solved to times larger than the inverse ultraviolet cutoff.

\begin{acknowledgments}
The author would like to thank K. Kajantie, L. McLerran and R. Fries for 
discussions that led to writing this paper
and R. Venugopalan for urging to actually write it up and
comments on the manuscript.
This manuscript has been authorized under Contract No. DE-AC02-98CH10886 
with the U.S. Department of Energy.
\end{acknowledgments}

\bibliographystyle{h-physrev4mod}
\bibliography{spires}

\begin{thebibliography}{10}

\bibitem{McLerran:1994ni}
L.~D. McLerran and R.~Venugopalan,
\newblock Phys. Rev. {\bf D49}, 2233 (1994), [arXiv:hep-ph/9309289].

\bibitem{McLerran:1994ka}
L.~D. McLerran and R.~Venugopalan,
\newblock Phys. Rev. {\bf D49}, 3352 (1994), [arXiv:hep-ph/9311205].

\bibitem{McLerran:1994vd}
L.~D. McLerran and R.~Venugopalan,
\newblock Phys. Rev. {\bf D50}, 2225 (1994), [arXiv:hep-ph/9402335].

\bibitem{Krasnitz:1998ns}
A.~Krasnitz and R.~Venugopalan,
\newblock Nucl. Phys. {\bf B557}, 237 (1999), [arXiv:hep-ph/9809433].

\bibitem{Krasnitz:1999wc}
A.~Krasnitz and R.~Venugopalan,
\newblock Phys. Rev. Lett. {\bf 84}, 4309 (2000), [arXiv:hep-ph/9909203].

\bibitem{Krasnitz:2000gz}
A.~Krasnitz and R.~Venugopalan,
\newblock Phys. Rev. Lett. {\bf 86}, 1717 (2001), [arXiv:hep-ph/0007108].

\bibitem{Krasnitz:2001qu}
A.~Krasnitz, Y.~Nara and R.~Venugopalan,
\newblock Phys. Rev. Lett. {\bf 87}, 192302 (2001), [arXiv:hep-ph/0108092].

\bibitem{Lappi:2003bi}
T.~Lappi,
\newblock Phys. Rev. {\bf C67}, 054903 (2003), [arXiv:hep-ph/0303076].

\bibitem{Lappi:2004sf}
T.~Lappi,
\newblock Phys. Rev. {\bf C70}, 054905 (2004), [arXiv:hep-ph/0409328].

\bibitem{Lappi:2006fp}
T.~Lappi and L.~McLerran,
\newblock Nucl. Phys. {\bf A772}, 200 (2006), [arXiv:hep-ph/0602189].

\bibitem{Fries:2006pv}
R.~J. Fries, J.~I. Kapusta and Y.~Li,
\newblock arXiv:nucl-th/0604054.

\bibitem{Gelis:2003vh}
F.~Gelis and R.~Venugopalan,
\newblock Phys. Rev. {\bf D69}, 014019 (2004), [arXiv:hep-ph/0310090].

\bibitem{Gelis:2004jp}
F.~Gelis, K.~Kajantie and T.~Lappi,
\newblock Phys. Rev. {\bf C71}, 024904 (2005), [arXiv:hep-ph/0409058].

\bibitem{Lappi:2006nx}
T.~Lappi,
\newblock arXiv:hep-ph/0606090.

\bibitem{Gelis:2005pb}
F.~Gelis, K.~Kajantie and T.~Lappi,
\newblock Phys. Rev. Lett. {\bf 96}, 032304 (2006), [arXiv:hep-ph/0508229].

\bibitem{Fujii:2005rm}
H.~Fujii, F.~Gelis and R.~Venugopalan,
\newblock Eur. Phys. J. {\bf C43}, 139 (2005), [arXiv:hep-ph/0502204].

\bibitem{Fujii:2005vj}
H.~Fujii, F.~Gelis and R.~Venugopalan,
\newblock Phys. Rev. Lett. {\bf 95}, 162002 (2005), [arXiv:hep-ph/0504047].

\bibitem{Kharzeev:2005iz}
D.~Kharzeev and K.~Tuchin,
\newblock Nucl. Phys. {\bf A753}, 316 (2005), [arXiv:hep-ph/0501234].

\bibitem{Kharzeev:2006zm}
D.~Kharzeev, E.~Levin and K.~Tuchin,
\newblock arXiv:hep-ph/0602063.

\bibitem{Kharzeev:2001ev}
D.~Kharzeev, A.~Krasnitz and R.~Venugopalan,
\newblock Phys. Lett. {\bf B545}, 298 (2002), [arXiv:hep-ph/0109253].

\bibitem{Kharzeev:2004ey}
D.~Kharzeev,
\newblock Phys. Lett. {\bf B633}, 260 (2006), [arXiv:hep-ph/0406125].

\bibitem{Hirano:2004rs}
T.~Hirano and Y.~Nara,
\newblock Nucl. Phys. {\bf A743}, 305 (2004), [arXiv:nucl-th/0404039].

\bibitem{Hirano:2004rsb}
T.~Hirano and Y.~Nara,
\newblock J. Phys. {\bf G30}, S1139 (2004), [arXiv:nucl-th/0403029].

\bibitem{Kovner:1995ja}
A.~Kovner, L.~D. McLerran and H.~Weigert,
\newblock Phys. Rev. {\bf D52}, 6231 (1995), [arXiv:hep-ph/9502289].

\bibitem{Kovner:1995ts}
A.~Kovner, L.~D. McLerran and H.~Weigert,
\newblock Phys. Rev. {\bf D52}, 3809 (1995), [arXiv:hep-ph/9505320].

\bibitem{Gyulassy:1997vt}
M.~Gyulassy and L.~D. McLerran,
\newblock Phys. Rev. {\bf C56}, 2219 (1997), [arXiv:nucl-th/9704034].

\bibitem{Kovchegov:1997ke}
Y.~V. Kovchegov and D.~H. Rischke,
\newblock Phys. Rev. {\bf C56}, 1084 (1997), [arXiv:hep-ph/9704201].

\bibitem{Iancu:2000hn}
E.~Iancu, A.~Leonidov and L.~D. McLerran,
\newblock Nucl. Phys. {\bf A692}, 583 (2001), [arXiv:hep-ph/0011241].

\bibitem{Kharzeev:2003wz}
D.~Kharzeev, Y.~V. Kovchegov and K.~Tuchin,
\newblock Phys. Rev. {\bf D68}, 094013 (2003), [arXiv:hep-ph/0307037].

\bibitem{Blaizot:2004wu}
J.~P. Blaizot, F.~Gelis and R.~Venugopalan,
\newblock Nucl. Phys. {\bf A743}, 13 (2004), [arXiv:hep-ph/0402256].

\bibitem{Gelis:2006tb}
F.~Gelis, A.~M. Stasto and R.~Venugopalan,
\newblock arXiv:hep-ph/0605087.

\bibitem{Jalilian-Marian:1997xn}
J.~Jalilian-Marian, A.~Kovner, L.~D. McLerran and H.~Weigert,
\newblock Phys. Rev. {\bf D55}, 5414 (1997), [arXiv:hep-ph/9606337].

\bibitem{Kovchegov:1996ty}
Y.~V. Kovchegov,
\newblock Phys. Rev. {\bf D54}, 5463 (1996), [arXiv:hep-ph/9605446].

\bibitem{Kovchegov:1997pc}
Y.~V. Kovchegov,
\newblock Phys. Rev. {\bf D55}, 5445 (1997), [arXiv:hep-ph/9701229].

\bibitem{Krasnitz:2002mn}
A.~Krasnitz, Y.~Nara and R.~Venugopalan,
\newblock Nucl. Phys. {\bf A717}, 268 (2003), [arXiv:hep-ph/0209269].

\bibitem{Kovchegov:2005ss}
Y.~V. Kovchegov,
\newblock Nucl. Phys. {\bf A762}, 298 (2005), [arXiv:hep-ph/0503038].

\bibitem{Kovchegov:2005kn}
Y.~V. Kovchegov,
\newblock Nucl. Phys. {\bf A764}, 476 (2006), [arXiv:hep-ph/0507134].

\bibitem{Gunion:1981qs}
J.~F. Gunion and G.~Bertsch,
\newblock Phys. Rev. {\bf D25}, 746 (1982).

\bibitem{Krasnitz:2003jw}
A.~Krasnitz, Y.~Nara and R.~Venugopalan,
\newblock Nucl. Phys. {\bf A727}, 427 (2003), [arXiv:hep-ph/0305112].

\bibitem{Romatschke:2005pm}
P.~Romatschke and R.~Venugopalan,
\newblock Phys. Rev. Lett. {\bf 96}, 062302 (2006), [arXiv:hep-ph/0510121].

\bibitem{Romatschke:2006nk}
P.~Romatschke and R.~Venugopalan,
\newblock Phys. Rev. {\bf D74}, 045011 (2006), [arXiv:hep-ph/0605045].

\bibitem{Golec-Biernat:1998js}
K.~Golec-Biernat and M.~Wusthoff,
\newblock Phys. Rev. {\bf D59}, 014017 (1999), [arXiv:hep-ph/9807513].

\bibitem{Golec-Biernat:1999qd}
K.~Golec-Biernat and M.~Wusthoff,
\newblock Phys. Rev. {\bf D60}, 114023 (1999), [arXiv:hep-ph/9903358].

\bibitem{Stasto:2000er}
A.~M. Stasto, K.~Golec-Biernat and J.~Kwiecinski,
\newblock Phys. Rev. Lett. {\bf 86}, 596 (2001), [arXiv:hep-ph/0007192].

\bibitem{Freund:2002ux}
A.~Freund, K.~Rummukainen, H.~Weigert and A.~Schafer,
\newblock Phys. Rev. Lett. {\bf 90}, 222002 (2003), [arXiv:hep-ph/0210139].

\bibitem{Kharzeev:2004if}
D.~Kharzeev, E.~Levin and M.~Nardi,
\newblock Nucl. Phys. {\bf A747}, 609 (2005), [arXiv:hep-ph/0408050].

\bibitem{Iancu:2003xm}
E.~Iancu and R.~Venugopalan,
\newblock arXiv:hep-ph/0303204.

\bibitem{Weigert:2005us}
H.~Weigert,
\newblock Prog. Part. Nucl. Phys. {\bf 55}, 461 (2005), [arXiv:hep-ph/0501087].

\end{thebibliography}

\end{document}